\documentclass[showpacs,twocolumn,superscriptaddress,aps]{revtex4}
\usepackage{amssymb}
\usepackage{graphicx}
\usepackage{dcolumn}
\usepackage{bm}
\usepackage{amsmath,amssymb}
\usepackage{graphicx}
\usepackage{color}

\begin{document}

\title{The classical capacity for continuous variable teleportation channel}
\author{QIN Tao}
\email{taoqin@mail.ustc.edu.cn}
\thanks{Telephone:0551-3607635}
\affiliation{Department of Modern Physics, University of Science and Technology of China,
Hefei 230026, People's Republic of China}
\author{ZHAO Mei-Sheng}
\affiliation{Department of Modern Physics, University of Science and Technology of China,
Hefei 230026, People's Republic of China}
\author{ZHANG Yong-De}
\affiliation{CCAST (World Laboratory), P.O. Box 8730, Beijing 100080, People's Republic
of China}
\affiliation{Department of Modern Physics, University of Science and Technology of China,
Hefei 230026, People's Republic of China}
\date{\today }

\begin{abstract}
The process of quantum teleportation can be considered as a quantum channel.
The exact classical capacity of the continuous variable teleportation
channel is given. Also, the channel fidelity is derived. Consequently, the
properties of the continuous variable quantum teleportation are discussed
and interesting results are obtained. Hopefully the method here can also be
applied to the cases for dense coding and swapping, which are crucial
quantum information protocols as well.
\end{abstract}

\maketitle


Unknown quantum states can be transmitted between distant users by means of
classical communication aided with quantum entanglement, which was labeled
\textit{teleportation} by its authors \cite{tele}. The teleportation process
can be regarded as a quantum channel \cite{chuang,preskill,pati,bowen}. The
teleportation channel is of great significance by exploiting the intriguing
properties of quantum entanglement. This protocol was initially put forth in
the discrete case. Vaidman generalized the protocol for quantum states of an
infinite dimensional Hilbert space, i.e., the teleportation for continuous
variables \cite{vaidman}. The teleportation protocols have been
experimentally demonstrated both discrete and continuous cases \cite%
{finite,infinite}.

The interest for continuous variable teleportation is motivated by the
following reasons:

(1). In continuous variable teleportation, although it is hard to generate
highly entangled states, the Bell measurement can be easily implemented with
the method of the homodyne detection \cite{infinite,zoller};

(2). As we will demonstrate, the capacity for the continuous variable
teleportation is proportional to the input mean photon number. Ideally,
there is no up bound to the channel capacity. Same conclusion holds in
classical continuous variable channels \cite{thomas}.

The theory of continuous variable quantum teleportation has been extensively
studied \cite{kimble,rmp,banjpa,ban2002,ban2003,ban2004,stefano}. The
formulae of continuous variable teleportation are presented and important
properties are discussed, especially the channel fidelity. These works are
quite stimulating.

Calculating the capacities of quantum channels has been a principal goal of
quantum information theory \cite{chuang,preskill}. However, the evaluations
of the capacities have been an intractable issue. Although many efforts have
been devoted to this endeavor \cite{capacity}, only a handful channels'
capacities have been solved. To our knowledge, the classical capacity for
continuous variable quantum teleportation channel is still obscure.

In this Letter, we address the problem of continuous variable quantum
teleportation channel by providing its classical capacity. Further, we give
the fidelity for this teleportation channel. And we present detailed
analysis of the properties of channel capacity and fidelity. Some
interesting results are obtained.

Quantum channels are completely positive, trace-preserving linear maps \cite%
{chuang,preskill}. Quantum teleportation can be regarded as such a map,
denoted as $\Lambda $. The map $\Lambda $ transforms an input state $\rho
_{in}$\ to an output state $\rho _{out}$. As for continuous variable
teleportation, it can be regarded as a bosonic channel. The messages are
encoded in the position and momentum quadratures of $\rho _{in}$, and then
decoded from $\rho _{out}$ via homodyne measurements.

The one shot channel capacity of $\Lambda $ is given by the
Holevo-Schumacher-Westmoreland (HSW) theorem \cite{holevo}%
\begin{equation}
\chi \left( \Lambda \right) \equiv \max_{\left\{ p_{i},\rho _{i}\right\} }
\left[ S(\Lambda (\sum_{i}p_{i}\rho _{i}))-\sum_{i}p_{i}S\left( \Lambda
\left( \rho _{i}\right) \right) \right]
\end{equation}%
where the maximum is taken over all ensembles $\left\{ p_{i},\rho
_{i}\right\} $ of possible input states $\rho _{i}$ to the channel. Here $%
p_{i}$ are the \textit{a priori} probability distributions, and $S\left(
\rho \right) =-Tr\left( \rho \log _{2}\rho \right) $ is the von Neumann
entropy. We declare here that the basis of the logarithm function is $2$ all
through the Letter. $\chi \left( \Lambda \right) $ is also called the Holevo
bound. So far as the purely lossy bosonic channel is concerned, it is proven
that a Gaussian mixture of coherent states, namely, a thermal state,
achieves the Holevo bound \cite{macchi,qin}.

In teleportation, Alice the sender and Bob the receiver share an arbitrary
bipartite quantum state $\rho _{AB}$ beforehand. When Alice is supplied with
an arbitrary unknown state $\rho _{in}$, she performs the joint measurement
of position and momentum for $\rho _{in}$ and the part of the bipartite
system in $\rho _{AB}$. Then Alice tells Bob the measurement results over a
classical channel. Obtaining the results, Bob employs the unitary
transformation to the other part of the $\rho _{AB}$ system to gain the
final output state $\rho _{out}$\ \cite{tele}.

Assume Alice and Bob share a two-mode squeezed-vacuum state $\rho _{AB}$
through a noisy quantum channel. $\rho _{AB}$ is denoted as%
\begin{eqnarray*}
\rho _{AB} &=&S\left[ s\left( a_{1}^{+}b_{1}^{+}-a_{1}b_{1}\right) \right]
\left\vert 0^{A}\right\rangle \left\langle 0^{A}\right\vert \\
&&\otimes \left\vert 0^{B}\right\rangle \left\langle 0^{B}\right\vert S\left[
-s\left( a_{1}^{+}b_{1}^{+}-a_{1}b_{1}\right) \right]
\end{eqnarray*}%
where $a_{1}$ $\left( b_{1}\right) $ and $a_{1}^{+}$ $\left(
b_{1}^{+}\right) $ are the bosonic annihilation and creation operators for
the modes $A$ $(B)$, respectively. $s\in \left[ 0,\infty \right) $ is the
squeezing parameter, representing how highly entangled $\rho _{AB}$ is. The
modes $A$ and $B$ are assigned to Alice and Bob, respectively.

For simplicity, we suppose that the input quantum state $\rho _{in}$ of the
continuous variable quantum teleportation is the coherent state
\begin{equation}
\rho _{in}=\left\vert \alpha \right\rangle \left\langle \alpha \right\vert
\end{equation}%
then the output state $\rho _{out}$ is given as follows \cite{ban2004}%
\begin{equation}
\rho _{out}=\Lambda \left( \rho _{in}\right) =\int d^{2}\beta q\left( \beta
\right) D\left( \beta \right) \rho _{in}D^{+}\left( \beta \right)
\end{equation}%
where $d^{2}\beta =d\Re \left( \beta \right) d\Im \left( \beta \right) $
while $D\left( \beta \right) =e^{\beta a^{+}-\beta ^{\ast }a}$ denotes the
displacement operator. Here $a^{+}\left( a\right) $ is the creation
(annihilation) operator of the input coherent state.\ The noise in this
continuous variable teleportation channel is presented by $q\left( \beta
\right) =\frac{1}{\pi \overline{n}_{s}}e^{-\frac{\left\vert \beta
\right\vert ^{2}}{\overline{n}_{s}}}$. Here $\overline{n}_{s}=2\left[ 1-T%
\left[ 1-\exp \left( -2s\right) \right] \right] $ ,which is the noise
variance, with $T\in \left[ 0,1\right] $ being the channel transmission
coefficient of the noisy channel \cite{ban2004}. When $T=1$, the
teleportation channel is noiseless. The teleportation channel $\Lambda $
then randomly displaces an input coherent state according to a Gaussian
distribution, which results in a thermal state.

According to the energy conservation law, there is an energy constraint to
the continuous variable teleportation channel \cite{macchi}
\begin{equation*}
\sum_{i}p_{i}Tr\left( \rho _{i}a^{+}a\right) \leq \overline{n}
\end{equation*}

To begin with, the integral of $\rho _{out}$ is derived as%
\begin{eqnarray}
\rho _{out} &=&\int d^{2}\beta q\left( \beta \right) D\left( \beta \right)
\left\vert \alpha \right\rangle \left\langle \alpha \right\vert D^{+}\left(
\beta \right)  \notag \\
&=&\int d^{2}\beta q\left( \beta \right) D\left( \beta \right) :e^{-\left(
a^{+}-\alpha ^{\ast }\right) \left( a-\alpha \right) }:D^{+}\left( \beta
\right)  \notag \\
&=&\int d^{2}\beta q\left( \beta \right) :e^{-\left( a^{+}-\alpha ^{\ast
}-\beta ^{\ast }\right) \left( a-\alpha -\beta \right) }:  \notag \\
&=&\frac{1}{1+\overline{n}_{s}}:e^{-\frac{1}{1+\overline{n}_{s}}\left(
a^{+}-\alpha ^{\ast }\right) \left( a-\alpha \right) }:  \notag \\
&=&\frac{1}{1+\overline{n}_{s}}D\left( \alpha \right) :e^{-\frac{1}{1+%
\overline{n}_{s}}a^{+}a}:D^{+}\left( \alpha \right)  \notag \\
&=&\frac{1}{1+\overline{n}_{s}}e^{\ln \frac{\overline{n}_{s}}{1+\overline{n}%
_{s}}\left( a^{+}-\alpha ^{\ast }\right) \left( a-\alpha \right) }
\end{eqnarray}

In order to calculate the channel capacity, we have to derive the thermal
state $\overline{\rho }$%
\begin{eqnarray*}
\overline{\rho } &=&\int d^{2}\alpha q\left( \alpha \right) \rho _{out} \\
&=&\int d^{2}\alpha \frac{1}{\pi \overline{n}}e^{-\frac{\left\vert \alpha
\right\vert ^{2}}{\overline{n}}}\rho _{out} \\
&=&\int d^{2}\alpha \frac{1}{\pi \overline{n}}e^{-\frac{\left\vert \alpha
\right\vert ^{2}}{\overline{n}}}\frac{1}{1+\overline{n}_{s}}:e^{-\frac{1}{1+%
\overline{n}_{s}}\left( a^{+}-\alpha ^{\ast }\right) \left( a-\alpha \right)
}: \\
&=&\frac{1}{\left( 1+\overline{n}_{s}\right) \overline{n}}\int \frac{%
d^{2}\alpha }{\pi }:e^{-\frac{1}{1+\overline{n}_{s}}\left( a^{+}-\alpha
^{\ast }\right) \left( a-\alpha \right) -\frac{1}{\overline{n}}\alpha ^{\ast
}\alpha }: \\
&=&\frac{1}{1+\overline{n}_{s}+\overline{n}}:e^{-\frac{1}{1+\overline{n}_{s}+%
\overline{n}}a^{+}a}: \\
&=&\frac{1}{1+\overline{n}_{s}+\overline{n}}e^{\frac{\overline{n}_{s}+%
\overline{n}}{1+\overline{n}_{s}+\overline{n}}a^{+}a}
\end{eqnarray*}

Hence, due to Eq. (1) the channel capacity is
\begin{equation}
\chi \left( \Lambda \right) =S\left( \overline{\rho }\right) -\int
d^{2}\alpha q\left( \alpha \right) S\left( \rho _{out}\right)
\end{equation}

Straightforward calculation shows
\begin{eqnarray}
S\left( \rho _{out}\right) &=&-Tr\left( \rho _{out}\log \rho _{out}\right)
\notag \\
&=&\log \left( 1+\overline{n}_{s}\right) Tr\left( \rho _{out}\right)  \notag
\\
&&-\log \left( \frac{\overline{n}_{s}}{1+\overline{n}_{s}}\right) Tr\left(
\rho _{out}\left( a^{+}-\alpha ^{\ast }\right) \left( a-\alpha \right)
\right)  \notag \\
&=&\log \left( 1+\overline{n}_{s}\right) -\overline{n}_{s}\log \left( \frac{%
\overline{n}_{s}}{1+\overline{n}_{s}}\right)  \notag \\
&=&\left( \overline{n}_{s}+1\right) \log _{2}\left( \overline{n}%
_{s}+1\right) -\overline{n}_{s}\log _{2}\overline{n}_{s}
\end{eqnarray}%
and%
\begin{eqnarray}
S\left( \overline{\rho }\right) &=&-Tr\left( \overline{\rho }\log \overline{%
\rho }\right)  \notag \\
&=&\log \left( 1+\overline{n}_{s}+\overline{n}\right)  \notag \\
&&-\log \left( \frac{\overline{n}_{s}+\overline{n}}{1+\overline{n}_{s}+%
\overline{n}}\right) Tr\left( \overline{\rho }a^{+}a\right)  \notag \\
&=&\log \left( \overline{n}+\overline{n}_{s}+1\right)  \notag \\
&&-\left( \overline{n}_{s}+\overline{n}\right) \log \left( \frac{\overline{n}%
_{s}+\overline{n}}{1+\overline{n}_{s}+\overline{n}}\right)  \notag \\
&=&\left( \overline{n}+\overline{n}_{s}+1\right) \log \left( \overline{n}+%
\overline{n}_{s}+1\right)  \notag \\
&&-\left( \overline{n}+\overline{n}_{s}\right) \log \left( \overline{n}+%
\overline{n}_{s}\right)
\end{eqnarray}

Substituting Eqs. (6) and (7) into Eq. (5), we obtain the channel capacity
\begin{eqnarray}
\chi \left( \Lambda \right) &=&\left( \overline{n}+\overline{n}_{s}+1\right)
\log \left( \overline{n}+\overline{n}_{s}+1\right)  \notag \\
&&-\left( \overline{n}+\overline{n}_{s}\right) \log \left( \overline{n}+%
\overline{n}_{s}\right)  \notag \\
&&-\left( \overline{n}_{s}+1\right) \log \left( \overline{n}_{s}+1\right)
\notag \\
&&+\overline{n}_{s}\log \overline{n}_{s}
\end{eqnarray}

For a vivid picture, the channel capacity is plotted in Figs. 1\&2, under
the circumstances of $\overline{n}=0.2$ and $\overline{n}=0.8$, respectively.

\begin{figure}[h]
\begin{center}
\includegraphics[width=0.4\textwidth]{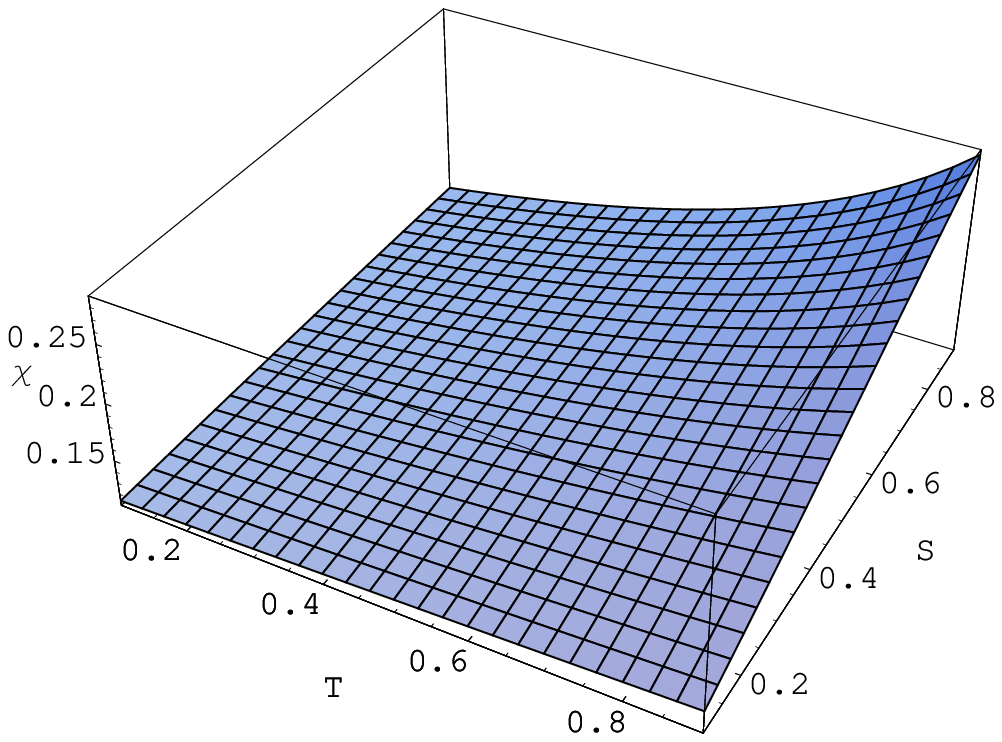}
\end{center}
\caption{The capacity when $\overline{n}=0.2$.}
\end{figure}

\begin{figure}[h]
\begin{center}
\includegraphics[width=0.4\textwidth]{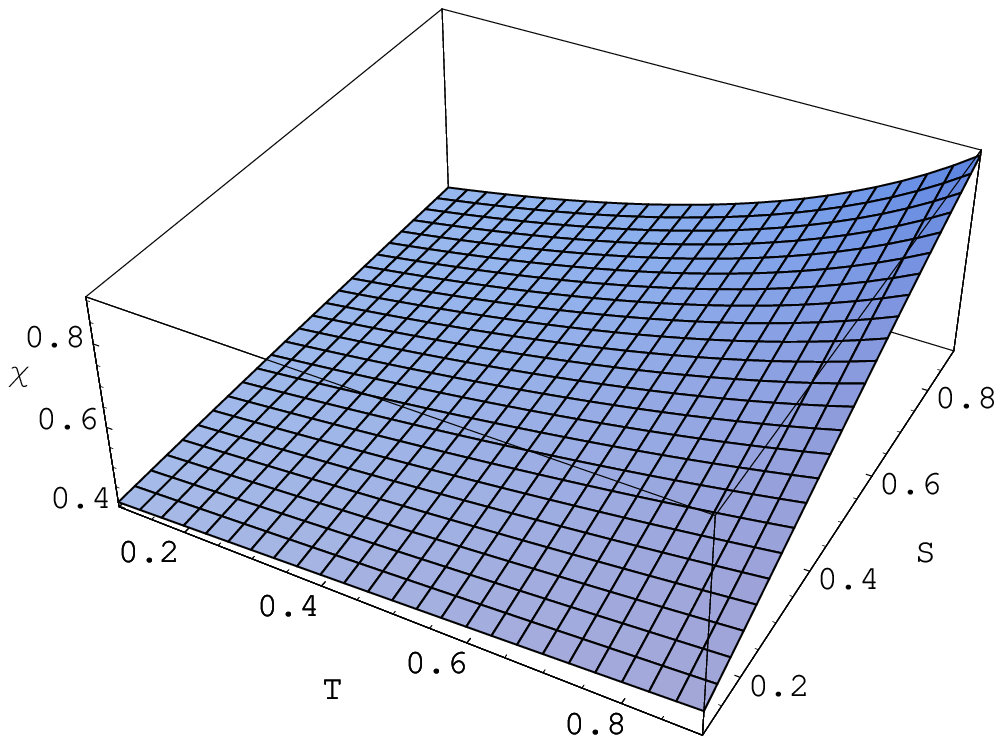}
\end{center}
\caption{The capacity when $\overline{n}=0.8$.}
\end{figure}

Examining these diagrams in Figs. 1\&2, we can see that the channel capacity
always increases with the input signal $\overline{n}$, $T$ and $s$. This is
comprehensive, since the noise intensity decreases with larger $T$ and $s$.
Meanwhile, $s$ represents how highly entangled $\rho _{AB}$ is. We observe
that larger $s$ gives rise to larger channel capacity. So it can be inferred
that to gain high channel capacity, the more entangled $\rho _{AB}$ is, the
better.

In addition, the fidelity measures the similarity between the input state
and the output state. And from the view of quantum information theory, it
evaluates how well the channel preserves the transmitted information \cite%
{chuang,preskill}. Thus the fidelity is an important characteristic of
quantum channels and a nontrivial quantity to study.

For the input state $\left\vert \alpha \right\rangle \left\langle \alpha
\right\vert $,\ the channel fidelity $F$ can be derived as%
\begin{eqnarray}
F &\equiv &\left\langle \alpha \right\vert \rho _{out}\left\vert \alpha
\right\rangle   \notag \\
&=&\left\langle \alpha \right\vert \frac{1}{1+\overline{n}_{s}}:e^{-\left(
\frac{1}{1+\overline{n}_{s}}+1\right) \left( a^{+}-\alpha ^{\ast }\right)
\left( a-\alpha \right) }:\left\vert \alpha \right\rangle   \notag \\
&=&\frac{1}{1+\overline{n}_{s}}\times e^{-\left( \frac{1}{1+\overline{n}_{s}}%
+1\right) \left( \alpha ^{\ast }-\alpha ^{\ast }\right) \left( \alpha
-\alpha \right) }  \notag \\
&=&\frac{1}{1+\overline{n}_{s}}
\end{eqnarray}

For a better understanding, the channel fidelity is plotted as in Fig. 3.

\begin{figure}[h]
\begin{center}
\includegraphics[width=0.4\textwidth]{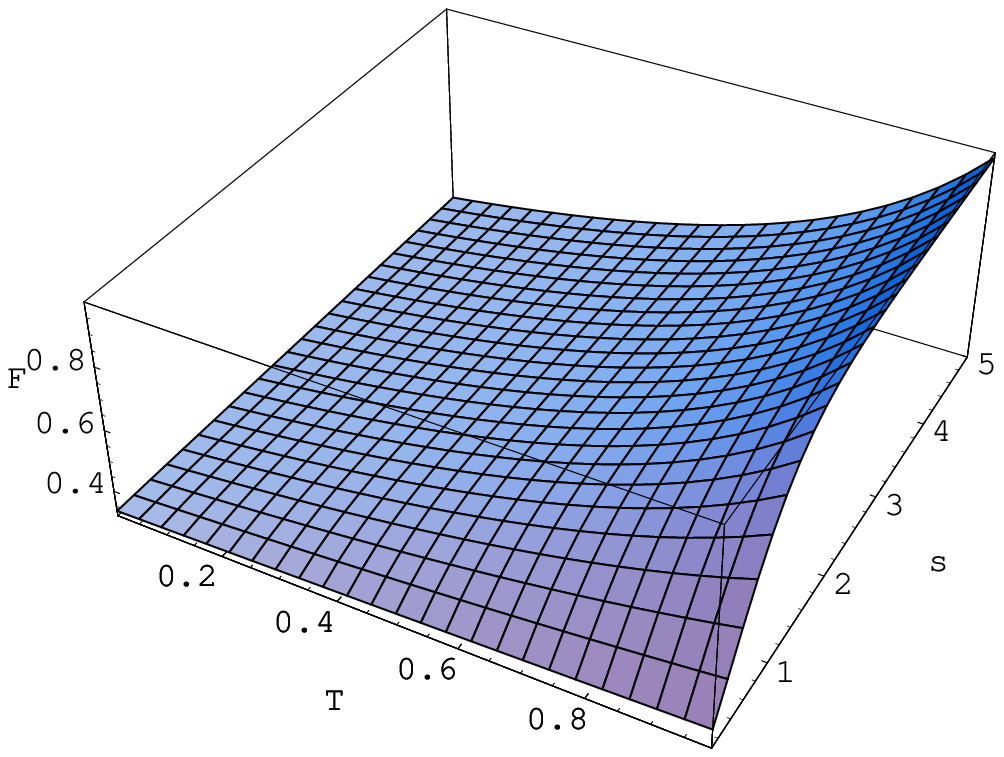}
\end{center}
\caption{The channel fidelity $F$.}
\end{figure}

Obviously the channel fidelity $F$ decreases with noise $\overline{n}_{s}$.
This is intuitive: The quantum state being transmitted in the channel will
become more distorted when the channel is more noisy. Specifically, the
channel fidelity increases with $T$ and $s$. As $s$ represents how highly
entangled $\rho _{AB}$ is, we can see that the more entangled $\rho _{AB}$
is, the larger the channel fidelity $F$ is.

On the other hand, the channel fidelity here is independent of input signal $%
\overline{n}$, viz., the mean photon number. This is not always the case. If
the bosonic channel is lossy, the channel fidelity decreases with the input
signal $\overline{n}$ \cite{qin}. Therefore, it can be concluded that the
channel fidelity is independent of input signal when the channel is not
lossy.

In conclusion, we employ the method in Refs. \cite{macchi,qin} to examine
the properties of continuous variable quantum teleportation. The exact
classical capacity of the continuous variable teleportation is given. We
find that the channel capacity $\chi $ increases with mean photon number of
the input state, and decreases with the growing intensity of the noise. The
channel capacity is closely related to the entangled state preshared by
Alice and Bob. The more highly entangled $\rho _{AB}$ is, the larger the
channel capacity is.

Furthermore, we give the exact channel fidelity in the case when the input
state is the coherent state. As expected, the fidelity drops off when the
channel becomes more noisy. And surprisingly the channel fidelity is
independent of the input mean photon number. The fidelity share the similar
properties of the channel capacity. The more entangled $\rho_{AB} $ is, the
larger the channel fidelity is.

The scenario we study here can be extended to more complex cases when
different input states other than coherent state, such as the squeezed
coherent states are considered. Hopefully more interesting outcomes can be
obtained.

Noticeably, the method to derive for channel capacity and fidelity in this
Letter can be applied to the cases for continuous variable quantum dense
coding \cite{dense,kimbledense} and continuous variable quantum swapping
\cite{swapping}. Both continuous variable dense coding and swapping can be
treated as quantum bosonic channels and they share resembled formulations as
continuous variable teleportation does. By making use of the same method in
this Letter, similar results will be expected.

Teleportation plays an important role in quantum information theory. We hope
our results can stimulate more interest on this subject.


\end{document}